\documentclass[pra,aps,preprint,byrevtex,showpacs]{revtex4}

\usepackage{epsfig}
\usepackage{amsmath}

\newcommand {\ket}[2] {| #1 \rangle_{#2}}
\newcommand {\eqn}[1] {Eq.~(\ref{#1})}
\newcommand {\half} {\frac{1}{2}}
\newcommand {\fig}[1] {Fig.~\ref{#1}}
\newcommand {\figwidth} {80mm}

\newcommand {\Sec}[1] {Sec.~\ref{#1}}

\newcommand {\ain} {\hat{a}_{\mathrm{in}}}

\newcommand {\an} {\hat{a}_n}
\newcommand {\bOne} {\hat{b}_1}
\newcommand {\bTwo} {\hat{b}_2}

\newcommand {\vac} {\hat{v}}

\newcommand {\Xplus} {\hat{X}^+}
\newcommand {\Xminus} {\hat{X}^-}

\newcommand {\Vplus} {V^+}

\newcommand {\Vminus} {V^-}

\newcommand {\etal} {\emph{et~al.}}

\newcommand {\ovAvFid} {\bar{\mathcal{F}}}

\begin{document}

\title{Optimal cloning for finite distributions of coherent states}

\author{P.~T.~Cochrane}
\email{cochrane@physics.uq.edu.au}
\affiliation{Department of Physics, The University of Queensland,
St.~Lucia, Brisbane, Queensland 4072, Australia}

\author{T.~C.~Ralph}
\affiliation{Department of Physics, The University of Queensland,
St.~Lucia, Brisbane, Queensland 4072, Australia}

\author{A.~Doli\'{n}ska}
\affiliation{Quantum Optics Group, Department of Physics, Faculty of
Science, Australian National University, ACT 0200, Australia}

\date{\today}

\begin{abstract}
We derive optimal cloning limits for finite Gaussian distributions of
coherent states, and describe techniques for achieving them. We discuss
the relation of these limits to state estimation and the no-cloning
limit in teleportation. A qualitatively different cloning limit is
derived for a single-quadrature Gaussian quantum cloner.
\end{abstract}


\maketitle

\section{Introduction}

Creating exact copies of unknown quantum states chosen from a
non-orthogonal set is impossible, due to the no-cloning
theorem~\cite{Wootters:1982:1, Barnum:1996:1}.  However, it is still
possible to make approximate copies of quantum states.  The best
achievable quality of the copy depends on the dimensionality of the
state Hilbert space, as well as the distribution of states picked from
that space.  For example, for a uniform distribution of states picked
from a qubit space, the best average overlap (fidelity) of the clones
with the original is $\frac{5}{6}$~\cite{Buzek:1996:1}.  For a flat
distribution over an infinite dimensional space the limit is
$\frac{2}{3}$~\cite{Cerf:2000:1}.  This paper investigates the
experimentally relevant situation, in continuous variables, where the
distribution of input states to the cloner is a finite distribution,
picked from an infinite Hilbert space.

In contrast to its counterpart in the single particle
regime~\cite{Wootters:1982:1,Barnum:1996:1}, cloning of continuous
variables~\cite{Cerf:2000:1} has only been investigated over the last
few years.  Gaussian cloning machines are of immediate interest for
continuous variables as they represent the optimal way to clone a wide
class of experimentally accessible states; the Gaussian states,
including coherent and squeezed states.  They are so called because
they add Gaussian distributioned noise in the cloning process.

We derive quantum cloning limits for finite distributions of coherent
states, and we investigate a method to tailor the standard
implementation using a linear amplifier~\cite{Braunstein:2001:2} to
take advantage of the known input state distribution.  We also
describe the qualitatively different quantum cloning limits for
coherent states with a distribution in the magnitude of their
amplitudes but with known phase; we will refer to this as states ``on
a line''. We also show that a Gaussian quantum cloner utilising an optical
parametric oscillator, as opposed to a linear amplifier, is the
optimum approach in this case.

The paper is arranged as follows: we begin in the next section by
reviewing the standard cloning limit for coherent states.  In
\Sec{sec:restrictedGauss} we examine the cloning of finite width
Gaussian distributions of coherent states and comment on the
connection between this and optimal state estimation. As well we
investigate how the ``no-cloning limit'' in teleportation is modified
for finite distributions.  In \Sec{sec:singleQuadCloner} we consider
the case of cloning coherent states on a line, and we conclude in
\Sec{sec:concl}.

\section{The standard cloning limit}
\label{sec:stdCloneLimit}

\begin{figure}[h]
\centerline{\includegraphics[width=\figwidth]{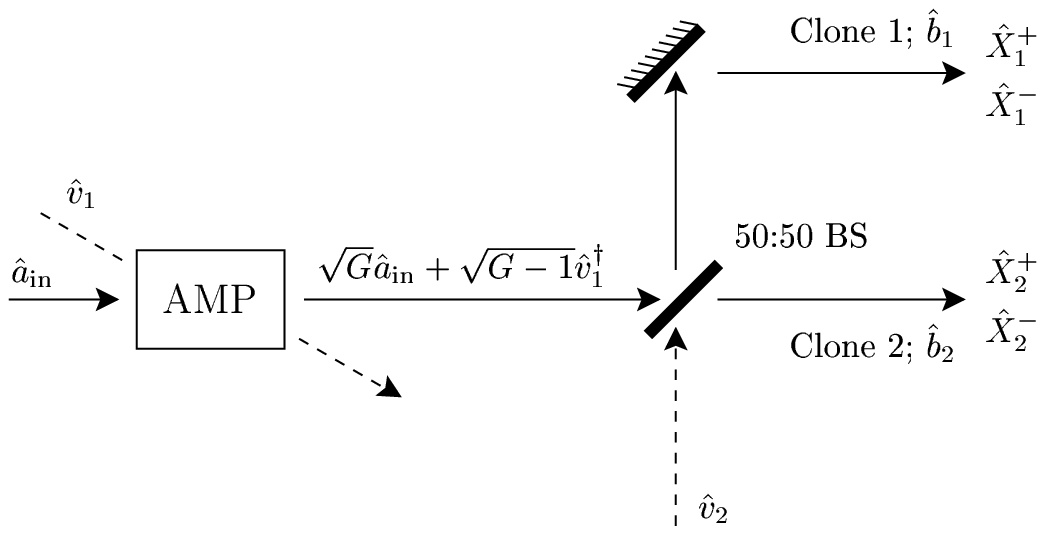}}
\caption{The optimum quantum Gaussian cloning machine, for complete
two quadrature copies.  The input field $\ain$ enters the amplifier
AMP and is amplified according to the gain $G$.  This field is then
incident onto a 50:50 beam splitter, the output from which forms the
two ``clones''.  Vacuum noise is added at both the amplifier and the
beam splitter.}
\label{AMP}
\end{figure}

An optimal (Gaussian) cloner for coherent states can be constructed
from a linear optical amplifier and a beam
splitter~\cite{Braunstein:2001:2}, as shown in \fig{AMP}.
The input field can be described by the annihilation operator
$\hat{a}_{in}$ and the initial coherent state $|\alpha \rangle$, where
$\alpha$ is a complex number representing the coherent amplitude of
the state. The Heisenberg evolution introduced by the linear amplifier
transforms this input field into the output field, $\hat{a}_{out} =
\sqrt{G} \hat{a}_{in} + \sqrt{G-1} \hat{v}_{1}^{\dagger}$. The field
is then divided on a 50:50 beamsplitter. The output modes are then
given by
\begin{align}
\bOne &= \frac{1}{\sqrt{2}} ( \sqrt{G} \ain + \sqrt{G-1}
\vac_1^{\dagger} + \vac_2 ),\\
\bTwo &= \frac{1}{\sqrt{2}} ( \sqrt{G} \ain + \sqrt{G-1}
\vac_1^{\dagger} - \vac_2 ).
\end{align}
Since both modes have the same amplitude and noise statistics, we need
only consider the quadrature amplitudes and variance of mode $1$.  The
quadrature amplitudes $\Xplus = \bOne + \bOne^\dagger$ and $\Xminus =
(\bOne - \bOne^\dagger)/i$ are
\begin{align}
\Xplus &= \frac{1}{\sqrt{2}} (\sqrt{G} \Xplus_{\ain} + \sqrt{G-1}
\Xplus_{\vac_{1}} + \Xplus_{\vac_{2}}),\\
\Xminus &= \frac{1}{\sqrt{2}} (\sqrt{G} \Xminus_{\ain} - \sqrt{G-1}
\Xminus_{\vac_{1}} + \Xminus_{\vac_{2}}),
\label{Xpm}
\end{align}
Assuming the input field is in a coherent state, the amplitude and
phase variances ($\Vplus$ and $\Vminus$ respectively) are given by
\begin{equation}
\Vplus = \Vminus = G.
\label{amp}
\end{equation}

The standard criterion for determining the efficacy of a given cloning
scheme is the fidelity of the input state with each of the clones. The
fidelity quantifies the overlap of the input state with the clone.  In
its simplest form, for two pure states, the fidelity is the modulus
squared of the inner product of the two states.  When the input is a
coherent state, the fidelity is given by the
expression~\cite{Furusawa:1998:1}
\begin{equation}
F = \frac{2}{\sqrt{(1+\Vplus)(1+\Vminus)}} \exp\left(
\frac{-2 (1-g)^2 |\alpha|^2}
{\sqrt{(1+\Vplus) (1+\Vminus)}}\right),
\label{eq:fidelitySimplified}
\end{equation}
where $g = \alpha_{clone}/\alpha$, is the amplitude gain of the
coherent amplitude of the clones ($\alpha_{clone} = \langle \bOne
\rangle = \langle \bTwo \rangle$) with respect to the coherent
amplitude of the input state ($\alpha = \langle \ain \rangle$).  Unit
gain ($g=1$) is the best cloning strategy when the input state is
completely arbitrary.  This is because the exponential dependence of
the fidelity on gain will dominate and lead to low fidelities for
large $\alpha_{in}$ unless the gain is exactly one. With unity gain
the fidelity becomes independent of the input state, and is thus only
a function of the output variances.

Picking unit gain by setting $G = 2$ and substituting \eqn{amp} into
\eqn{eq:fidelitySimplified} gives an average fidelity (defined by
$\ovAvFid = \int F(\alpha) P(\alpha) d^2\alpha$) of $\ovAvFid =
\frac{2}{3}$.  Since the fidelity does not depend on the amplitude of
the input state at unity gain we have $\ovAvFid = F \int P(\alpha)
d^2\alpha$, hence $\ovAvFid = F$.

The optimality of this result was proved by
Cerf~\etal~\cite{Cerf:2000:2} by considering the generalized
uncertainty principle for measurements~\cite{Arthurs:1965:1}.  When
applied to coherent states this principle requires that in any
symmetric, simultaneous measurement of the two quadrature amplitudes,
sufficient noise is added such that the signal to noise of the two
measurement results is reduced by at least a half over what would be
obtained by an ideal measurement of one or other of the
quadratures. This result implies that the minimum amount of noise that
can be added in the cloning process is just enough so that the
signal to noise of the quadratures of the clones (as would be found in
an ideal single quadrature measurement) is reduced to precisely one
half of that of the original state. This is just sufficient to prevent
the generalized uncertainty principle being violated by performing an
ideal measurement of, say, the amplitude quadrature of clone 1 and the
phase quadrature of clone 2.  Using \eqn{Xpm} it is easy to show that
the signal to noise ratios of the quadratures of each clone are equal
and given by $\mathrm{SNR}^{\pm}_{out} =
\frac{1}{2}\mathrm{SNR}^{\pm}_{in}$. To be more explicit, the input
field $\ain$ can be written as $\ain = \alpha + \an$. In this
representation the initial state is now the vacuum (giving rise to
quantum noise) and the coherent amplitude, $\alpha$ (now included
explicitly in the Heisenberg evolution), is considered the signal.
The output mode $\bOne$ can now be written as
\begin{equation}
\bOne = \frac{1}{\sqrt{2}} (\sqrt{G}(\alpha + \an) + \sqrt{G-1}\vac_1 
+ \vac_2).
\end{equation}
The signal to noise transfer ratio ($T = \mathrm{SNR}^{\pm}_{out}/
\mathrm{SNR}^{\pm}_{in}$) of either quadrature of either clone is
\begin{equation}
T = \frac{\half G}{\half (G + G-1 + 1)} = \half.
\label{T}
\end{equation}
Thus each clone has the minimum noise added to it allowed by quantum
mechanics, and thus is optimal.

So far we have assumed (as in all previous discussions of continuous
variable cloners) that the input state distribution is uniform over
all quadrature-phase space; the probability of seeing a given state at
the input of the cloner is the same for \emph{all} states.  However,
this implies an infinite distribution which, for practical reasons, is
not the case experimentally.  Therefore, in general, one has some
knowledge about the input state distribution. We now consider how this
information can be used to improve the output fidelity of the cloner
by tailoring the gain to the input state distribution.

\section{Cloning a Restricted Gaussian Distribution}
\label{sec:restrictedGauss}

Let us consider a two-dimensional Gaussian distributed coherent input
state distribution with mean zero and variance $\sigma^{2}$:
\begin{equation}
P(\alpha) = \frac{1}{2\pi\sigma^2}
\exp\left(\frac{-\alpha_x^2-\alpha_y^2}{2\sigma^2}\right),
\label{eq:twoDimGaussian}
\end{equation}
where $\alpha_x$ and $\alpha_y$ are the real and imaginary parts
respectively of the input coherent state. Such a distribution is
optimal for encoding information~\cite{Shannon:1948:1}
and is experimentally accessible.

Using this distribution, we can find the average fidelity by
integrating the fidelity for a given state $\ket{\alpha}{}$
[\eqn{eq:fidelitySimplified}] weighted by the probability of obtaining
that state, $P(\alpha)$, over all $\alpha$.  This is described
mathematically by
\begin{equation}
\ovAvFid = \int F(\alpha) P(\alpha) d^2\alpha.
\label{eq:ovAvFidDefn}
\end{equation}

We maximise this fidelity over the gain of the amplifier, $G$ to
obtain $\ovAvFid$ as a function only of the variance of the input
state distribution.  Knowing the distribution of input states can now
allow us to choose an appropriate amplifier gain to maximise the
cloning fidelity.  Since the minimum value of $G$ is 1, $\ovAvFid$ is
a piecewise continuous function of $\sigma$; the two pieces of the
function being joined at $\sigma^2 = \half + \frac{1}{\sqrt{2}}$.  The
average fidelity is given by
\begin{equation}
\ovAvFid = \left\{
\begin{aligned}
& \frac{4\sigma^2 + 2}{6\sigma^2 + 1},
&\sigma^2 \ge \half + \frac{1}{\sqrt{2}} \approx 1.2\\
& \frac{1}{(3 - 2\sqrt{2})\sigma^2 + 1},
&\sigma^2 \le \half + \frac{1}{\sqrt{2}} \approx 1.2.
\end{aligned}
\right.
\label{eq:ovAvFid}
\end{equation}
Since we maximise the average fidelity over the gain, $G$ is
implicitly a function of $\sigma$, and is given by
\begin{equation}
G = \frac{8 \sigma^4}{(2 \sigma^2 + 1)^2},
\end{equation}
when $\sigma^2 \ge \half + \frac{1}{\sqrt{2}}$, and by $G = 1$
otherwise.

\begin{figure}
\centerline{%
\epsfig{file=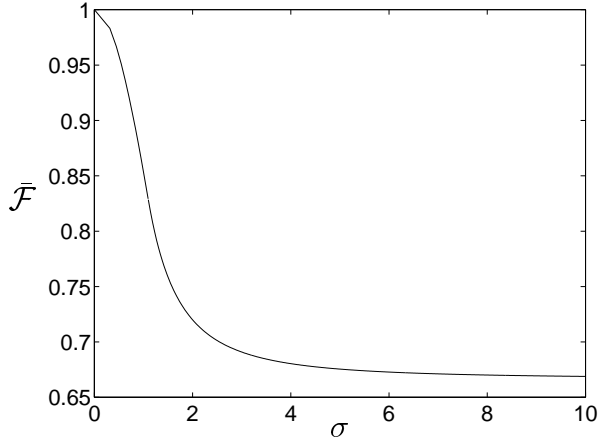,width=\figwidth}%
}
\caption{Average cloning fidelity $\ovAvFid$ (maximised over the amplifier 
gain) versus standard deviation of the
input state probability distribution $\sigma$.  The figure shows that for
large $\sigma$ the average fidelity is equal to the cloning limit of
$\ovAvFid=2/3$, and that as $\sigma$ decreases, the average fidelity
increases to unity, because the prior knowledge one has of the input state 
becomes perfect in this limit.}
\label{fig:FbarSigma}
\end{figure}

The average fidelity maximised over the amplifier gain is shown as a
function of $\sigma$ in \fig{fig:FbarSigma}.  
Notice that at large $\sigma$, $\ovAvFid(\sigma)$ is at the standard
cloning limit $\ovAvFid = 2/3$.  In other words, for sufficiently
broad input state distributions, the situation is equivalent to having
a completely arbitrary input state.  As $\sigma$ decreases, the
fidelity increases, because we now have better knowledge of the likely
value of the input state; approaching unit fidelity as $\sigma$ tends
to zero.  This is an intuitive result, since if $\sigma=0$ then the
input state distribution is a two-dimensional delta function in
quadrature-phase space and we know with certainty the value of the
input state (i.e. it is the vacuum) prior to cloning.

Notice that in \eqn{T} the minimum allowed noise added to the clones
does not depend on the gain of the amplifier $G$.  Our procedure of
tuning the gain to find the maximum fidelity has retained this
property and our clones are therefore optimal.

\subsection{State estimation}

We now consider the connection between cloning and optimal state
estimation.  Dual homodyne detection (or equivalently heterodyne
detection) is known to be the optimal technique for estimating the
amplitude of an unknown coherent state drawn from a Gaussian
ensemble~\cite{Arthurs:1965:1,Yuen:1973:1}.  For our setup, dual
homodyne detection of the input state would correspond to setting the
amplifier gain to $G=1$, and detecting the amplitude quadrature of one
of the output beams and the phase quadrature of the other. Given that
we know the standard deviation of the input state distribution
$\sigma$, it can be shown that the best estimate of the amplitude of
the input state is given by
\begin{equation}
\alpha^\prime =  \frac{1}{\sqrt{2}} \left( \frac{2 \sigma^2}{2 \sigma^2 + 1}
\right)
\left( \Xplus + i \Xminus \right),
\label{est}
\end{equation}
where $\Xplus$ and $\Xminus$ are the measured values of the amplitude
and phase quadratures respectively.  In the limit of broad
distributions ($\sigma \to \infty$) the best estimate is just given by
\begin{equation}
\alpha^\prime =  \frac{1}{\sqrt{2}} \left( \Xplus + i \Xminus \right).
\end{equation}
However, as the distribution narrows it is better to underestimate the
value of $\alpha$ in accordance with \eqn{est}. In the limit of
$\sigma \to 0$ we become certain that $\alpha$ is zero regardless of
the measurement outcome.

We have observed that the signal to noise transfer between the
original state and the clones is not changed by the choice of
amplifier gain, thus optimal state estimation must be possible using
the clones. Some insight into the physics of the particular choice of
amplifier gain which produces the optimal clones can be obtained by
noticing that for optimal clones, the best estimation of the original
state amplitude is determined by measuring the amplitude quadrature of
one, and the phase quadrature of the other, and then setting
\begin{equation}
\alpha^\prime =  \frac{1}{2} \left( \Xplus + i \Xminus \right).
\end{equation}
This is true for all distribution sizes down to the point where the
cloning amplifier gain, $G$, equals one. For even smaller
distributions we return to the dual homodyne formula. At such small
distribution sizes the quantum noise dominates.

\subsection{Teleportation and the no-cloning limit}

Teleportation is the entanglement assisted communication of quantum
states through a classical channel~\cite{Bennett:1993:1}.
Teleportation of continuous variables~\cite{Braunstein:1998:2} can be
achieved using entanglement of the form
\begin{equation}
\ket{\psi}{} = \sqrt{1 - \lambda^{2}}\sum_{n} \lambda^{n} \ket{n,n}{},
\end{equation}
which describes a two-mode squeezed vacuum.  The strength of the
entanglement is characterised by the parameter $\lambda$ which is
related to the squeezing, or noise reduction, of the quadrature
variable correlations of the modes.  Zero entanglement is
characterised by $\lambda = 0$, and maximum entanglement occurs when
$\lambda = 1$. Various ways to characterise the quality of the
teleportation process have been proposed
\cite{Braunstein:2001:1,Ralph:1998:1,Grosshans:2001:1}, and have been
used to describe experimental
demonstrations~\cite{Furusawa:1998:1,Bowen:2003:2}.

In terms of fidelity two distinct bounds have been identified for the
case of an infinite input distribution of unknown coherent states. The
classical limit is $\ovAvFid = 1/2$~\cite{Furusawa:1998:1}.
Fidelities higher than this value cannot be achieved in the absence of
entanglement, i.e. when $\lambda = 0$. The no-cloning limit on the
other hand requires that the teleported version of the input state is
demonstrably superior to that which could be possessed by anyone else.
This is not guaranteed unless the teleported state has an average
fidelity $\bar{\mathcal{F}} \ge
2/3$~\cite{Grosshans:2001:1}. Achieving this requires a particular
quality of entanglement, $\lambda \ge 1/3$, or more than $50\%$
squeezing. We now investigate how this no-cloning limit changes as a
function of the distribution of the input states.

\begin{figure}
\centerline{\epsfig{file=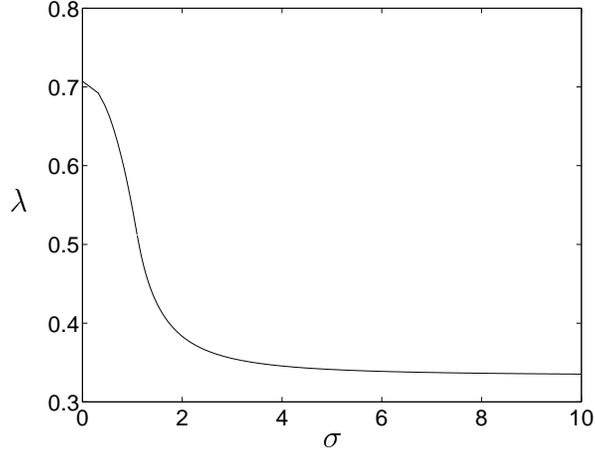,width=\figwidth}}
\caption{Entanglement required to reach the no-cloning limit as a
function of the input state distribution width. The squeezing
parameter $\lambda$ is plotted against the standard deviation,
$\sigma$, in dimensionless units.}
\label{fig:noclone}
\end{figure}

\begin{figure}
\centerline{%
\epsfig{file=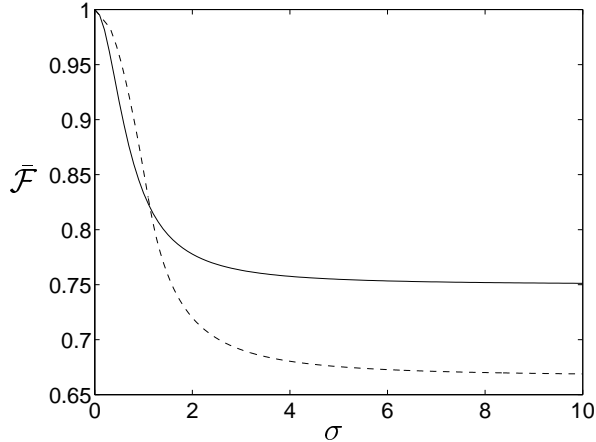,width=\figwidth}%
}
\caption{Teleportation fidelity $\bar{\mathcal{F}}$ (solid curve) and 
the no-cloning limit (dashed curve) as a function of distribution width,
$\sigma$; the level of squeezing is held constant at $\lambda = 0.5$.  
The teleportation fidelity beats the no-cloning limit for a large
range of distribution widths, but dips under for $\sigma \lesssim 1$.
The quantities presented are dimensionless.}
\label{fig:teleFidAndCloningLimit}
\end{figure}

In a previous paper~\cite{Cochrane:2003:1} two of us (PTC and TCR)
numerically optimised the average fidelity of continuous variable
teleportation of a finite Gaussian distribution of coherent input
states for various levels of entanglement.  An analytical expression
for this optimised average fidelity is given by
\begin{equation}
\bar{\mathcal{F}} = \frac{1 - 2(\lambda^2-1) \sigma^2}
{1 - 4(\lambda-1) \sigma^2},
\label{tele}
\end{equation}
where $\sigma$, as before, is the standard deviation of the input
state distribution. Using this result and that derived earlier for the
optimum cloning fidelity, [\eqn{eq:ovAvFid}], we can find the
squeezing ($\lambda$) required for the teleportation fidelity
[\eqn{tele}] to equal the no-cloning limit as a function of the
distribution width ($\sigma$).  This is given by
\begin{equation}
\lambda = \left\{
\begin{aligned}
& \frac{8\sigma^2 + 16\sigma^4 - 2\sqrt{2} \sqrt{\sigma^2
      (2\sigma^2-1)(1+2\sigma^2)^4}}
  {4\sigma^2 + 24\sigma^4},
&\sigma^2 \ge \half + \frac{1}{\sqrt{2}} \approx 1.2\\
& \frac{6 + 4\sqrt{2} - \sqrt{2}\sqrt{3 + 2\sqrt{2} + \sigma^2 +
      2\sigma^4}}
  {6 + 4\sqrt{2} + 2\sigma^2},
&\sigma^2 \le \half + \frac{1}{\sqrt{2}} \approx 1.2.
\end{aligned}
\right.
\end{equation}
The result is plotted in \fig{fig:noclone}.
The maximum squeezing parameter value $\lambda = 1/\sqrt{2}$ is
achieved at $\sigma = 0$.  This is the minimum amount of squeezing
required for teleportation to beat the no-cloning limit for all values
of $\sigma$.  Below this value it is possible for the teleportation
fidelity to be lower than the no-cloning limit---and for teleported
states not to be superior to that possessed by another party---for
some values of $\sigma$.  This is demonstrated in
\fig{fig:teleFidAndCloningLimit} where $\lambda = 0.5$.
The dashed curve is the no-cloning limit fidelity
and the solid curve is the teleportation fidelity as a function of
$\sigma$ for constant $\lambda = 0.5$.  Notice that for $\sigma \lesssim 1$
the teleportation fidelity drops just below the no-cloning limit.  Only with
$\lambda = 1/\sqrt{2}$ can one be sure that one will beat the no-cloning
limit.

Notice that in the limit of large $\sigma$ the teleportation fidelity
is higher than the no-cloning limit and that at $\sigma=0$ the
fidelity equals the no-cloning limit.  The lowest constant value
$\lambda$ can take and still equal the no-cloning limit at both
$\sigma \rightarrow \infty$ and $\sigma=0$ is $\lambda=1/3$,
corresponding to the quality of entanglement mentioned above.  At
lower squeezing parameter values the teleportation fidelity does not
achieve the no-cloning limit except for the trivial case of $\sigma=0$.

A somewhat surprising feature is that the quality of entanglement
required to reach the no-cloning limit actually increases as the width
of the distribution is decreased. This occurs because of the different
ways in which the teleporter and the cloner add noise at unity gain.
For example for a distribution with a standard deviation $\sigma = 3$,
an entanglement of $\lambda = 0.4$ is required, significantly higher
than the $\lambda = 1/3$ level of entanglement needed for an infinite
distribution.
\begin{figure}
\centerline{\epsfig{file=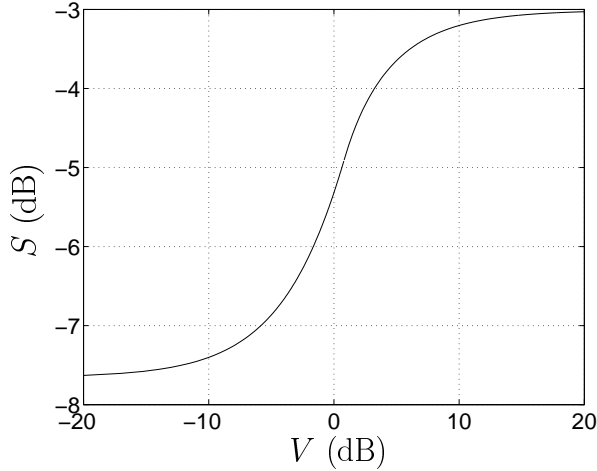,width=\figwidth}}
\caption{Entanglement required to reach the no-cloning limit as a
function of the input state distribution width, as in
\fig{fig:noclone}.  This figure plots the experimentally familiar
variables of squeezing ($S$) against the variance ($V$), in units of
decibels.}
\label{fig:noiseRed}
\end{figure}
\fig{fig:noiseRed} shows the same plot as \fig{fig:noclone} but using
the more experimentally familiar parameters of noise reduction (or
squeezing) of the entanglement,
\begin{equation}
S = \frac{(1-\lambda)^2}{1-\lambda^2},
\end{equation}
and the variance (or noise power) of the distribution
\begin{equation}
V = \sigma^{2},
\end{equation}
both plotted in decibels. These graphs show that the issue of the
no-cloning limit for teleportation is rather subtle when the realistic
situation of finite distributions of input states is taken into
account.

\section{The single quadrature cloning limit}
\label{sec:singleQuadCloner}

We now consider cloning of a rather different distribution of coherent
states; one in which all the coherent amplitudes have the same phase,
but have a broad distribution in the absolute value of their
amplitudes. Effectively the signals are encoded on only one
quadrature.  In \Sec{sec:stdCloneLimit} we discussed the optimum
cloning limit in terms of a restriction imposed by the uncertainty
principle between the two conjugate parameters being
copied~\cite{Cerf:2000:2}.  For information on a single quadrature,
where only one of the conjugate observables is being copied, it could
be easy to reach the na\"{\i}ve conclusion that there will not be such
a cloning limit.  However, quantum mechanics still places a
restriction on the fidelity of such clones because coherent states,
even when restricted to a line in phase space, are non-orthogonal.  If
the input states carry information on one quadrature only, a different
cloning process is required.  A new cloning limit, with a much higher
fidelity emerges in such a scenario and is now discussed.

\begin{figure}[h]
\begin{center}
\includegraphics[width=\figwidth]{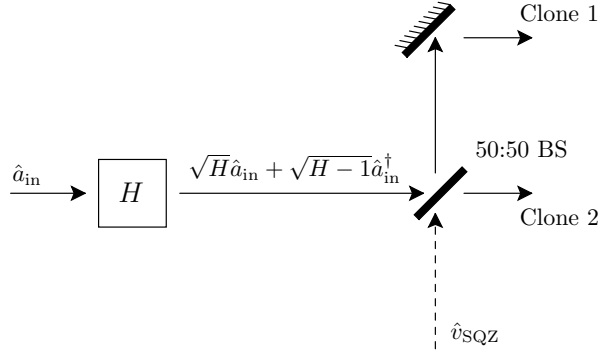}
\caption{The single quadrature quantum cloner. $H$ is the OPO gain.}
\label{OPOg}
\end{center}
\end{figure}

Consider the single quadrature Gaussian cloner shown in \fig{OPOg}.
It consists of a phase sensitive amplifier, an optical parametric
oscillator (OPO) set to amplify in the real direction \cite{Ben,
Walls:1995:1, PK, Lam:1997:1, OPO2}, followed by a 50:50 beam splitter
also made phase sensitive by the injection of squeezed vacuum noise at
the dark port, $\vac_{SQZ}$.  The output of the OPO is given by
$\sqrt{H} \hat{a}_{in} + \sqrt{H-1} \hat{a}_{in}^{\dagger}$ where $H$
is the parametric gain.  After passing through the beam splitter, the
variances of the output quadratures are:
\begin{equation}
V_{Clone1}^{\pm}\ =\ V_{Clone2}^{\pm}\ =\ \frac{V_{a}^{\pm}}{2}\
\left(\sqrt{H-1}\ \pm \sqrt{H}\right)^{2}\ +\ \frac{V_{\vac}^{\pm}}{2}.
\end{equation}

Suppose that the input distribution is now described by the
non-symmetric Gaussian distribution
\begin{equation}
P(\alpha) = \frac{1}{2\pi\sigma^2}
\exp\left(\frac{-\alpha_x^2}{2\sigma_{x}^2}
+ \frac{-\alpha_y^2}{2\sigma_{y}^2}\right).
\label{eq:twoDimGaussianAS}
\end{equation}
We assume $\sigma_{y} \ll 1$, restricting the coherent states to the
real axis. For simplicity we assume that the distribution ``along the
line'' is sufficiently broad, $\sigma_{x} \gg 1$, such that fidelity
will be optimized by unity gain operation. Unity gain is achieved by
setting the gain to $H = \frac{9}{8}$.  A minimum uncertainty state is
assumed for the squeezed input noise, i.e $V_{\vac}^{+}\ =
\frac{1}{V_{\vac}^{-}}$.  This gives a fidelity of:
\begin{equation}
F \rightarrow \frac{2}{\sqrt{\left(\frac{5}{4}+\frac{1}{2
V_{\vac}^{+}}\right)\left(2\ + \frac{V_{\vac}^{+}}{2}\right)}},
\label{Fidel2}
\end{equation}
which reaches a global maximum when the beam splitter input phase is
quadrature squeezed such that $V_{\vac}^{-} = \sqrt{\frac{5}{8}}$, (so
$V_{\vac}^{+} = \sqrt{\frac{8}{5}}$).  The maximum fidelity of the
clone is then $\frac{4}{9}(\sqrt{10}-1)$. An equivalent result can be
achieved by not injecting squeezed vacuum at the BS but instead
inserting two independent OPOs in each beam splitter output arm.

The $\mathrm{SNR}^{+}$ of the individual amplitude quadrature clones
is found to be:
\begin{equation}
\mathrm{SNR}_{out}^{+}\ =\frac{2\ \mathrm{SNR}_{in}^{+}}{(2\ + V_{\nu}^{+})}.
\end{equation}
With optimised fidelity at $V_{\vac}^{+}\ = \sqrt{\frac{5}{8}}$, the
above expression reduces to $\mathrm{SNR}_{out}^{+}\ =\frac{5\
\mathrm{SNR}_{in}^{+}}{5\ + \sqrt{10}} \approx$ 0.6125
$\mathrm{SNR}_{in}^{+}$.  This is a greater value than the single
quadrature average SNR and lies outside the classical regime.

Unlike the symmetric case discussed in \Sec{sec:stdCloneLimit}, the
single quadrature clones are entangled.  The SNR of the summed
amplitude quadratures of the clones, $X^{+}_{SUM}\ =\ X^{+}_{Clone1}\
+X^{+}_{Clone2}$, is independent of the vacuum squeezing parameter and
gives $\mathrm{SNR}_{out}^{+}\ =1$. This indicates that overall no
noise has been added in the cloning process.  We also note that
the noise outputs of the phase and amplitude quadratures are very
close to the minimum uncertainty product.

It seems likely that this is the optimum fidelity attainable for
coherent states on a line. The approach is analogous to the standard
cloner setup, and it is hard to imagine how the phase sensitive
amplification and phase sensitive beam splitter combination could be
improved upon given that overall no noise is added.  However, the
argument is not as straightforward as for a symmetric cloner because
the maximization of the fidelity is non-trivial, depending upon the
phase and strength of the squeezing injected at the beam splitter. An
extensive search of the parameter space revealed no better result,
and we conjecture that the fidelity is optimal.

\section{Conclusion}
\label{sec:concl}

We have shown how to tailor the Gaussian quantum cloner to optimally
clone unknown coherent states picked from finite symmetric Gaussian
distributions. Operating the cloner at a particular level
below unity gain maximises the cloning fidelity for such
distributions. This maximum fidelity increases monotonically as a
function of the distribution width from $2/3$ in the limit of very
broad distributions, to $1$ in the limit of very narrow
distributions. We discussed the relationship between this optimal gain
and state estimation, and have shown that the no-cloning limit for
teleportation of coherent states changes in a non-trivial way as a
function of the width of the input state distribution.

We have also demonstrated the existence of a qualitatively different
cloning limit for coherent states on a line, and have described a
machine which clones these states with a fidelity of
$F=\frac{4}{9}(\sqrt{10}-1)$.

\begin{acknowledgments}
The authors would like to thank P.~K.~Lam for helpful discussions.
This work was supported by the Australian Research Council.  The
diagrams in this paper were produced with PyScript.
\texttt{http://pyscript.sourceforge.net}.
\end{acknowledgments}

%\bibliography{thesis,bibase}

\end{document}